\documentstyle[aps,prl,epsfig]{revtex}

\newcommand{\ket}[1]{\left|{#1}\right\rangle}

\newcommand{\innx}[3]{\left\langle{\vphantom{#3}#1}\right|
        {#2}\left|{\vphantom{#1}#3}\right\rangle}

\newcommand{\im}{{\rm i}}
\newcommand{\xp}{{\scriptscriptstyle +}}

\def\lint{\int\limits}
\begin{document}
\draft
\twocolumn[\hsize\textwidth\columnwidth\hsize\csname %
@twocolumnfalse\endcsname

\title{ Numerical study of the $E\otimes e$
Jahn-Teller polaron and bipolaron}
\author{ $^a$S. El Shawish,  $^{a,b}$J. Bon\v ca, $^c$Li-Chung Ku,
and  $^c$S. A. Trugman}
\address{$^a$J. Stefan Institute, 1000, Slovenia,
$^b$FMF, University of Ljubljana, 1000, Slovenia and\\
$^c$Theory Division,
Los Alamos National Laboratory, Los Alamos, NM  87545,}
\date{\today}
\maketitle
\begin{abstract}\widetext

The properties of the polaron and bipolaron are explored in the 1D
Jahn-Teller model with dynamical quantum phonons. 
The ground-state properties of the polaron and bipolaron are computed
using a recently developed variational method.
Dynamical properties of the ground
state of a polaron are investigated by calculating the optical
conductivity $\sigma(\omega)$. Our numerical results suggest that the
Jahn-Teller and Holstein polarons are similar. However, in the
strong-coupling regime qualitative differences in $\sigma(\omega)$
between the two models are found and discussed.
The influence of the electron-phonon
coupling and the electrostatic repulsion on the bipolaron binding
energy, bipolaron masses, and correlation functions is investigated.

\end{abstract}

]

\narrowtext

\section{Introduction}
While research on Jahn-Teller (JT) electron phonon coupling spans
many decades, a renewed interest in this problem has been sparked by the
discovery of colossal magnetoresistance materials (CMR). In CMR's, it
is believed, the colossal magnetoresistance effect is a consequence of
the interplay between the double exchange mechanism and the lattice effect
\cite{millis,zhao,quijada,mareo}. Furthermore, recent studies of alkali doped
fullerenes $A_nC_{60}$ indicate that the Jahn-Teller distortion together
with strong Coulomb repulsion could be responsible for unusual
electronic properties of these materials
\cite{brouet,ole,auer,manini,capone1}.

Recent advances in computing capabilities have stimulated development
of various numerical techniques such as: exact diagonalization
techniques (ED) on small clusters \cite{marsiglio,wellein,mello}, ED
on infinite clusters \cite{bonca,lichung}, quantum Monte Carlo
calculations (QMC) \cite{raedt,kornilovitch}, variational methods
\cite{romero}, and density matrix renormalization group techniques
(DMRG) \cite{white}, that have provided valuable results for the
Holstein model (HM) in one and more spatial dimensions. 
The most efficient methods provide energies for
the Holstein polaron problem that are accurate
up to 21 digits in the thermodynamic limit \cite{lichung}.
In addition to static quantities, dynamic properties
such as spectral functions and
optical conductivity of the Holstein model have recently been
studied on small lattice clusters
\cite{fehske1,capone,fehske,alex,zhang}.  However, except in the small
polaron regime, such calculations are subject to pronounced
finite-size effects. In contrast to the HM, there has been much less
numerical research devoted to the Jahn-Teller model (JTM), mainly due
to the far larger Hilbert space that presents the main obstacle to
exact-diagonalization approaches. Most of the recent numerical
calculations of the JTM that take into account the full quantum
mechanical nature of the problem consider only one atom
\cite{ole,auer}. A recent path-integral Quantum Monte Carlo approach,
developed by Kornilovitch, has proven powerful in computing
ground state and spectral properties like the density of states of the
Jahn-Teller problem \cite{korni}. Calculation of dynamic properties
by that method would require analytic continuation from imaginary time,
which is very sensitive to statistical noise.

The scope of the present work is to compare
the static properties of the JTM and Holstein model (HM) for
the case of one and two electrons. In addition, we investigate dynamic
properties of the Holstein and Jahn-Teller polaron by calculating the
optical conductivity. In this case we devote equal attention to the
Holstein and Jahn-Teller models since dynamical properties of the
Holstein model calculated using an infinite-lattice variational space
have not been published  elsewhere.  We use a recently developed
numerical technique \cite{bonca,bonca1} to study the JTM for the case
of one and two electrons on the 1D infinite lattice.  Our main goal is
to find the numerically exact solution of the JTM in the thermodynamic
limit. The variational method that we use \cite{bonca,bonca1} is
defined on an infinite lattice and is not subject to finite-size
effects.  A standard Lanczos method is used to find ground and excited
states with the selected Hilbert space. The method allows calculation
of physical properties at any wavevector $k$. In the intermediate
coupling regime where it is most accurate, it provides results that
are variational in the thermodynamic limit and gives energies accurate
to 21 digits for the Holstein polaron, up to 7 digits for the
Holstein bound bipolaron, and 6 digits for the Jahn-Teller
polaron.  While there are no boundary finite-size effects, there are
nevertheless finite-variational-space effects due to a constraint that
only a finite separation between the electron and the surrounding
phonons is allowed for the polaron, and between two electrons and
the polaron cloud for the bipolaron. Nevertheless, results may be
fairly accurate even in the two-electron case when electrons are
bound into a bipolaron, since then electron-electron
correlation functions decrease exponentially for large separations.

\section{Model}
The simplest, so called $E\otimes e$ Jahn-Teller  model, consists
of two degenerate electron orbitals and two degenerate phonon modes,
\begin{eqnarray} \label{en_JTB}
   H &=& -t\sum_{i,o,s}\left(c_{i\xp 1,o,s}^{\xp} c_{i,o,s} + {\rm
         H.c.}\right) + \omega_0\sum_i\left(a_i^{\xp} a_i + b_i^{\xp}
         b_i\right) + \nonumber \\ &&
         +\ g\omega_0 \sum_{i,s}\left[(n_{i,1,s} -
         n_{i,2,s})(a_i^{\xp} + a_i)+
         \right. \nonumber \\ && \left.
         (c_{i,1,s}^{\xp} c_{i,2,s} +
         c_{i,2,s}^{\xp} c_{i,1,s})(b_i^{\xp} + b_i)\right]+ \nonumber
         \\ &&
         +\ U_1\sum_{i,o}n_{i,o,\uparrow} n_{i,o,\downarrow} + \
         \nonumber \\ &&
         U_2\sum_i\left(n_{i,1,\uparrow} n_{i,2,\downarrow}
         +n_{i,2,\uparrow} n_{i,1,\downarrow}\right),
\label{jth}
\end{eqnarray}
where $c_{i,o,s}^{\xp}$ creates an electron on site $i$,
orbital $o$ ($o=1,2$ ) and spin $s$, and $a_i^{\xp}$ ($b_i^{\xp}$) creates
dispersionless phonons of type $a$ ($b$) on site $i$. While the
phonon of type $a$ couples to the electron density, the mode $b$
mediates hopping between electronic orbitals that are orthogonal in
the absence of phonon $b$.  The parameters of the model are the
intersite hopping matrix element $t$, dimensionless electron-phonon
coupling strength $g$, and optical phonon frequency $\omega_0$. In the
bipolaron case we consider electrons with opposite spin.
The last two terms in Eq. (\ref{jth}) represent the on-site
same-orbital ($U_1$) and different-orbital ($U_2$) Coulomb repulsion.
For simplicity we have chosen $U_1=U_2=U$.
%

Most of the calculations will be compared  to a simpler Holstein
model
\begin{eqnarray}
   H &=& -t\sum_{i,s}\left(c_{i\xp 1,s}^{\xp} c_{i,s} + {\rm
       H.c.}\right)+\omega_0\sum_i a_i^{\xp} a_i \nonumber \\
     &+&g\omega_0 \sum_{i,s}
       n_{i,s}(a_i^{\xp} + a_i) + U \sum_i n_{i,\uparrow}
       n_{i,\downarrow}.
\label{eq:hm}
\end{eqnarray}

While the basic principles of the method have been explained elsewhere
\cite{bonca,bonca1} we here present only a brief explanation of how
the variational space is constructed for the case of two electrons in
the JTM.  Basis states for the many-body Hilbert space can be written
$\ket{\phi} = \ket{j_1,o_1,j_2,o_2;\ldots n_i,n_{i+1}\ldots
m_i,m_{i+1}\ldots}$, where the $j_1,j_2$ and $o_1,o_2$ are first and
second electron site and orbital indices respectively, and there are
$n_i$ phonons of type $a$ and $m_i$ phonons of type $b$ on site $i$. A
variational subspace is constructed beginning with an initial state where
both electrons are on the same site and orbital with no phonons, and operating
repeatedly ($N_h$ times) with the off-diagonal pieces ($t$ and $g$) of
the Hamiltonian, Eq. (\ref{en_JTB}).  The wavefunction is then written
in a translation invariant form.
To achieve high accuracy in the strong coupling limit
where there are many excited phonon quanta, we have constructed towers
of $N_{ad}$ ($N_{ad}=30$ for JTM and $N_{ad}=200$ for HM) additional
phonon excitations (for the JT case: one tower for each phonon type)
that were located on the electron site and on the first neighbor site
to the left {\bf or} to the right of the electron position. Such
towers play a crucial role in achieving convergence in the small
polaron regime.

\section{Jahn-Teller polaron}

\subsection{Static correlation functions}

\begin{figure}[]
\begin{center}
\epsfig{file=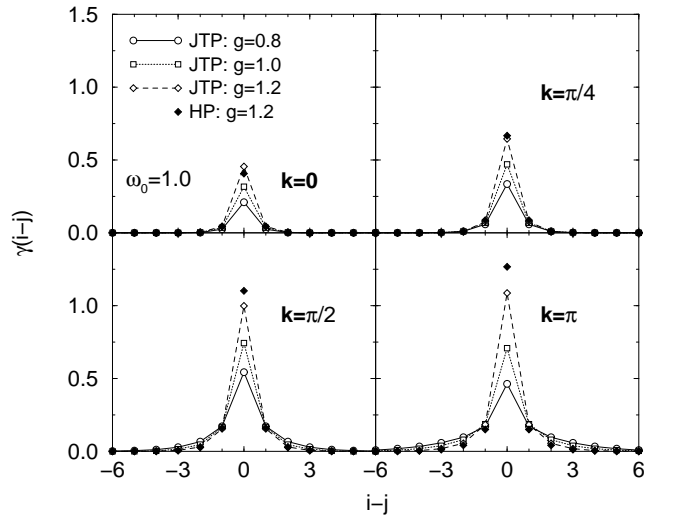,width=70mm,angle=-90}
\end{center}
\caption{ $\gamma(i-j)$ vs. $(i-j)$ for the JTP calculated at three
different coupling strengths and $\omega_0=t=1$ (open symbols). Results
of the HP are presented as black diamonds. }
\label{fig:gamma1}
\end{figure}

We start by presenting results of the JT polaron (JTP). To
investigate the shape of the JTP, we have computed various
static correlation functions. We first present the correlation
function for the distribution of the number of excited phonons in the
vicinity of the electron
\begin{equation}
   \gamma(i-j) = \innx{\psi_k}{n_i (a_j^{\xp}a_j + b_j^{\xp}b_j)}{\psi_k},
\label{eq:gamma}
\end{equation}
where $ n_i = n_{i,1} + n_{i,2} $ and $\ket{\psi_k}$ is the polaron
wave function at momentum $k$. Numerical results presented in
Fig.~(\ref{fig:gamma1}) show $\gamma(i-j)$ calculated at various coupling
strengths and different wavenumbers. When comparing results at different
$k$ it becomes evident that the size of the polaron at $k=\pi$ is larger
than that at $k=0$.  As the coupling strength increases, $\gamma(0)$
increases while $\gamma(i-j\not=0)$ does not change substantially at
$k=0$ while for larger values of $k$, $\gamma(i-j\not=0)$
slightly decreases.  We should stress that due to the
the symmetry of the pseudospin rotation \cite{takada},
$\innx{\psi_k}{n_ia_j^{\xp}a_j}{\psi_k} =
\innx{\psi_k}{n_ib_j^{\xp}b_j}{\psi_k}$. In Fig.~(\ref{fig:gamma1}) we
also compare JTP with the Holstein polaron (HP), where $\gamma$
is calculated at $g=1.2$. 
While  differences at small $k$ 
are very subtle, they become more pronounced at  larger $k$ where
$\gamma(0)$ of the HP is larger than that of the JTP.

\begin{figure}[]
\begin{center}
\epsfig{file=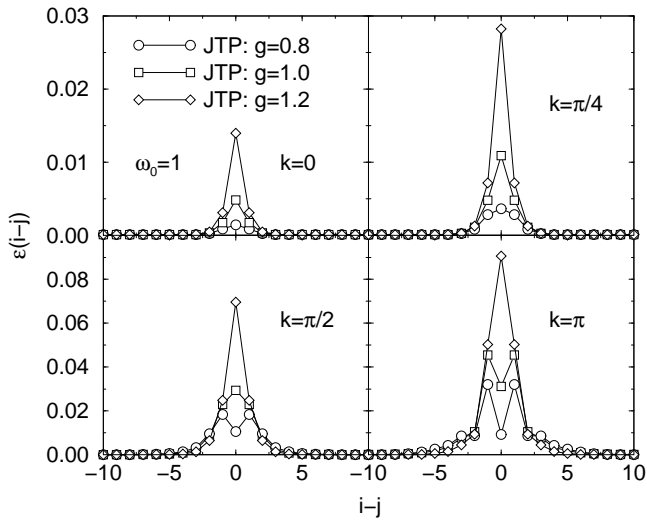,width=70mm,angle=-90}
\end{center}
\caption{$\varepsilon(i-j)$ vs. $(i-j)$ calculated at three different
coupling strengths and $\omega_0=t=1$. Note that vertical scales are
different for $k=0,\pi/4$ and $k=\pi/2,\pi$.}
\label{fig:epsi}
\end{figure}

Next we present the phonon-phonon correlation function of the JTP
\begin{equation}
   \varepsilon(i-j) = \innx{\psi_k}{a_i^{\xp}a_i b_j^{\xp}b_j}{\psi_k},
\label{eq:eps}
\end{equation}
shown in Fig.~\ref{fig:epsi}. As expected, phonons of two different
types are only weakly correlated. 
At small values of $k$ the electron
mediated phonon-phonon interaction is always attractive (it has a
maximum at $(i-j)=0$), however, at larger values of $k$ and in the
weak to intermediate coupling regime it becomes repulsive
($\varepsilon(i-j)$ peaks at $(i-j)=\pm 1$).
In contrast, the expectation for the same phonon type,
$\innx{\psi_k}{a_i^{\xp}a_i a_j^{\xp}a_j}{\psi_k}$,
always peaks at $(i-j)=0$ (not shown).

\begin{figure}[]
\begin{center}
\epsfig{file=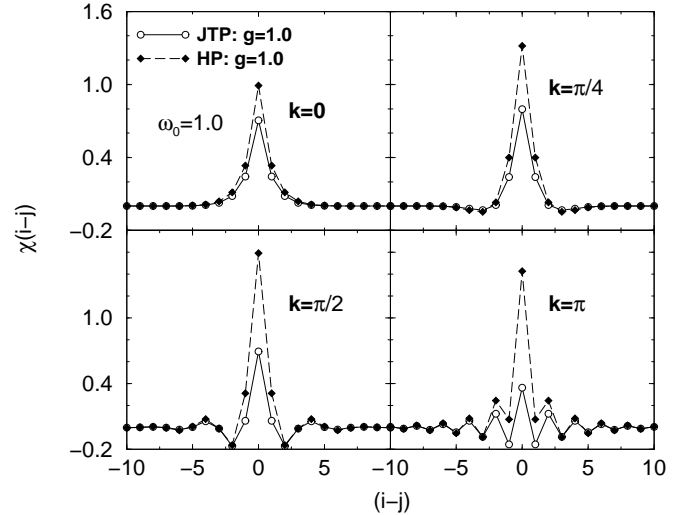,width=70mm,angle=-90}
\end{center}
\caption{$\chi(i-j)$ vs. $(i-j)$ for JTP and HP, $t=1$.}
\label{fig:chi}
\end{figure}

The static correlation function between the electron position and the
oscillator displacement is defined as
\begin{equation}
   \chi(i-j) = \innx{\psi_k}{n_i (a_j^{\xp} + a_j)}{\psi_k}.
\label{eq:chi}
\end{equation}
We should stress that when using properly symmetrized ground-state
wavefunction in the pseudo-spin space,  we numerically obtain  the
following equality: $ \innx{\psi_k}{n_i (b_j^{\xp} + b_j)}{\psi_k}=
\innx{\psi_k}{n_i (a_j^{\xp} + a_j)}{\psi_k}$.  In
Fig.~(\ref{fig:chi}) we present $\chi(i-j)$ in the intermediate
coupling regime, {\it i.e.} $g=1$.  At $k=0$, where the group velocity
is zero, the deformation around the electron position is limited to
only a few lattice sites, $\chi(i-j)$ is always positive and seems to
be exponentially decaying.  At finite but small $k=\pi/4$, the local
deformation around the electron increases in amplitude.  We also
notice a ringing effect (oscillating deformation) as a consequence of
a finite velocity at $k\not= 0$. At $k=\pi/2$ the ringing effect is
strongly enhanced while the spatial extent of the deformation
increases in comparison to $k=0$. The range of deformation reaches its
maximum at $k=\pi$ while at the same time the lattice displacement on
the electron site decreases in comparison to smaller values of $k$. In
comparison with HP (also presented in Fig.~(\ref{fig:chi})), the main
difference is in $\gamma(0)$ which in the case of the JTP diminishes
strongly with increasing $k$.

%

\begin{figure}[]
\begin{center}
\epsfig{file=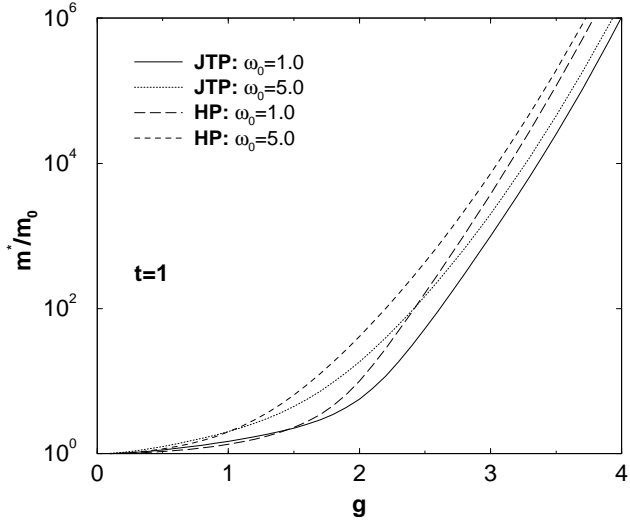,width=70mm,angle=-90}
\end{center}
\caption{Effective mass of the JTP and HP vs. $g$ for $t=1$,
$\omega_0=1$ and $5.0$, log scale.}
\label{fig:mass1a}
\end{figure}

An important advantage of defining the Hilbert space on an infinite
lattice is in the ability  to calculate the energy of the system $E$
at arbitrary value of $k$. This makes our method extremely efficient
for computing the effective mass of the polaron
\begin{equation}
  {m^*}^{-1} =  
  {\partial ^ 2 E(k) \over {\partial k ^ 2}}\vert_{k=0}.
\label{eq:mass}
\end{equation}
The second derivative is evaluated by finite differences in the
neighborhood of $k = 0 + dk$. Comparison between the JTP and HP
effective mass is presented in Fig.~(\ref{fig:mass1a}). In accordance
with the Kornilovitch results \cite{korni}, the effective mass of the HP
polaron is smaller than the JTP  in the weak to intermediate
coupling regime. With increasing coupling the JTP becomes lighter
than the HP. From Fig.~(\ref{fig:mass1a}) it seems as if the
effective mass of the JTP and HP  displays similar behavior in the
strong coupling limit. 
%
%

\begin{figure}[]
\begin{center}
\epsfig{file=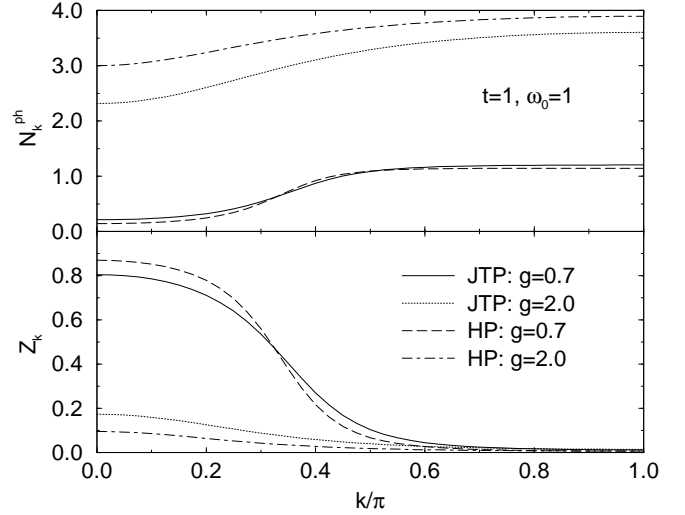,width=70mm,angle=-90}
\end{center}
\caption{$N_{k}^{ph}$ (top) and $Z_k$ (bottom) vs. $k/\pi$ for two
different coupling strengths. }
\label{fig:zknk}
\end{figure}

The evolution  of a polaron state with changing $k$
can be inspected by calculating
the quasiparticle weight and the mean phonon number $N_k^{ph}$:
\begin{eqnarray}
Z_k &=& \vert\langle\psi_k\vert c_k^\dagger\vert0\rangle\vert^2,
\label{eq:zk}\\
N^{ph}_k&=&\sum_i  \langle\psi_k\vert a_i^\dagger a_i+b_i^\dagger b_i\vert\psi_k
\rangle,
\label{eq:nk}
\end{eqnarray}
where $c_k^\dagger\vert0\rangle$ represents a state with an electron and no
excited phonons and $\vert \psi_k\rangle$ is the solution of the model
at finite $k$. Note that the difference between $Z_{k=0}$ and the
inverse effective mass Eq.~(\ref{eq:mass}) comes from $k-$ dependence
of the polaron self-energy $\Sigma(k,\omega)$ \cite{lichung}. At
finite $k$, $Z_k$ measures the electronic character of the polaronic
wavefunction. In Fig.~(\ref{fig:zknk}) we present $Z_k$ and
$N_k^{ph}$ for the HP and JTP calculated at two different
coupling strengths $g=0.7$ and $g=2.0$. Consistent with results for
the effective mass in Fig.~(\ref{fig:mass1a}), $Z_k$ of the JTP
is slightly larger than the HP at $k=0$, $g=2$. With increasing $k$ we
observe a smooth crossover from the predominantly electronic
character (large $Z_k$) towards the phononic character (small $Z_k$)
of a polaron.  This crossover is sharper in the HP case as $Z_k$
of the JTP intersects $Z_k$ of the HP close to $k=0.3
\pi$ $(g=0.7)$. This intersection is also reflected in $N_k^{ph}$ even though
$N_k^{ph}$ does not differ substantially between the two models.
In the case of  large coupling, $g=2.0$, $Z_k$ for the JTP  is
larger  than $Z_k$ of HP which is consistent with the behavior of the
effective mass in the strong coupling regime. Moreover, $N_k^{ph}$ of
the JTP is substantially smaller than $N_k^{ph}$ of HP.

\subsection{Optical conductivity}

Applying standard linear-response theory, the
real part of the conductivity takes the form (in the limit $T=0$),
\begin{equation}
   \Re\sigma(\omega)=D \delta(\omega) + \sigma^{reg}(\omega > 0),
\end{equation}
where $D$ is the Drude weight at $\omega=0$ and $\sigma^{reg}$ is the
finite-frequency response given by
\begin{eqnarray}
   \sigma^{reg}(\omega)=\pi e^2 \sum_n\frac{\left|
   \innx{\psi_0}{J}{\psi_n}\right|^2}{E_n-E_0}\delta(\omega+E_0-E_n),
\label{eq:sigom}
\end{eqnarray}
where all states $\vert \psi_n\rangle$ for $n=1,2,\dots$ belong to
$k=0$ space.  It is important to stress that in Eq.~(\ref{eq:sigom}),
the current operator $J$ has the same form in the Jahn-Teller and
Holstein model,
\begin{equation}
   J=-\im et\sum_{i,o}\left(c_{i\xp 1,o}^{\xp} c_{i,o} - {\rm
   H.c.}\right).
\label{eq:curr}
\end{equation}
Therefore we should expect a similar behavior of
$\sigma^{reg}(\omega)$ in both models.  We next  introduce the
$\omega$-integrated spectral weight
\begin{equation}
   S^{reg}(\omega)=\lint_{0^{\xp}}^{\omega} \sigma^{reg}(\omega)d\omega
\label{eq:som}
\end{equation}
and arrive at the sum rule
\begin{eqnarray}
   S^{tot}=\frac{\pi e^2}{2} \langle -H_{kin}\rangle = S^{reg} + \pi e^2 t
   {m^*}^{-1},
\label{eq:sumrule}
\end{eqnarray}
where $S^{reg}=S^{reg}(\infty)$. The sum rule may be used to test the
numerics by calculating the effective mass by first using Eq.~(\ref{eq:mass})
and second, Eq.~(\ref{eq:sumrule}). In all cases presented the numerical
sum rules are satisfied to at least 6 digits.

While our numerical method, defined on the infinite lattice, enables
us to keep the numerical precision in the thermodynamic limit of
static correlation functions including $m^*, S^{tot}$ and $S^{reg}$ within
a linewidth, there are finite-size effects when presenting
finite-frequency quantities. 
The reason is that the obtained spectrum
is discrete and we are thus forced to use a small $\epsilon\sim 0.05$
to smooth the continued-fraction form of Eq.~\ref{eq:sigom}.  On the
other hand, the integrated spectral weight $S^{reg}(\omega)$ does not
depend on $\epsilon$ and is thus more reliable.

\begin{figure}[]
\begin{center}
\epsfig{file=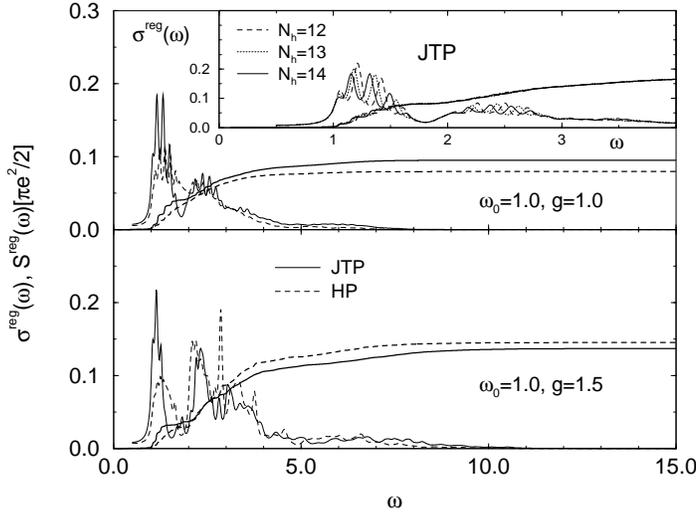,width=70mm,angle=-90}
\end{center}
\caption{$\sigma^{reg}(\omega)$ and $S^{reg}(\omega)$ vs. $\omega$ for
$g=1$ (top) and $g=1.5$ (bottom) for JTP (full lines) and HP (dashed
lines). Inset (top) shows $\sigma^{reg}(\omega)$ and
$S^{reg}(\omega)$ for the JTP calculated at three sizes of the Hilbert
space with $N_h=12,13,$ and $14$. Convergence of $S^{reg}(\omega)$ is
within the linewidth.}
\label{fig:sigma}
\end{figure}

We first present $\sigma^{reg}(\omega)$ and its integral
$S^{reg}(\omega)$ in Fig.~(\ref{fig:sigma}) for the JT and HM in the
weak to intermediate coupling regime. The precision of our results in
the thermodynamic limit can be tested in the weak coupling limit,
where the optical conductivity threshold is known to be at
$\omega=\omega_0$. In this case, other methods, defined on finite
systems, show pronounced finite-size effects due to discreteness in the
$k-space$ which are reflected in the threshold, larger than $\omega_0$
in the weak-coupling limit \cite{fehske1,capone,fehske}, and rather
well separated peaks corresponding to scattering of the initial $k=0$
electron state into finite-$k$ states.  The main signatures of
$\sigma^{reg}(\omega)$ at $g=1.0$ of both models (JT and HM) are: the
spectra are strongly asymmetric in frequency, which is a
characteristic of large polarons \cite{emin}.  Both spectra seem to
share the same threshold frequency $\omega_0$ \cite{lowfreq}. 
The first two broad
peaks are due to one and two phonon emission processes. Peaks are more
pronounced in the JT than in the HM case. With increasing coupling $g$,
$\sigma^{reg}(\omega)$ of both models shifts towards higher frequencies
and remains similar. The threshold frequency does not change significantly.

We feel obliged to discuss the constraint of the finite Hilbert space
effects on presented spectra in more detail. In the inset of
Fig.~(\ref{fig:sigma}) we present $\sigma^{reg}(\omega)$ for the JTP
calculated for three different sizes of the Hilbert space. We can
clearly see that broad features  (two broad peaks) in
$\sigma^{reg}(\omega)$ converge rather well, however, the number and
position of small peaks that compose  broad peaks change slightly with
increasing $N_h$. Excited states that contribute to
$\sigma^{reg}(\omega)$ in the total $k=0$ sector can be represented as
a polaron with the wavevector $k_p$ and excited phonon or phonons with
the total wavevector $k_{ph}=-k_p$.  Small peaks therefore represent
scattering of the polaron into different $k_p$ states with one or
multiple phonon emissions. Since in our variational space phonons are
allowed only $N_h$ steps away from the electron, there are only
discrete $k_p$'s allowed for such scattering.  In the inset of
Fig.~(\ref{fig:sigma}) we present also the integrated conductivity
$S^{reg}(\omega)$ for the same sizes of the Hilbert space. The
convergence in this case is excellent.

\begin{figure}[]
\begin{center}
\epsfig{file=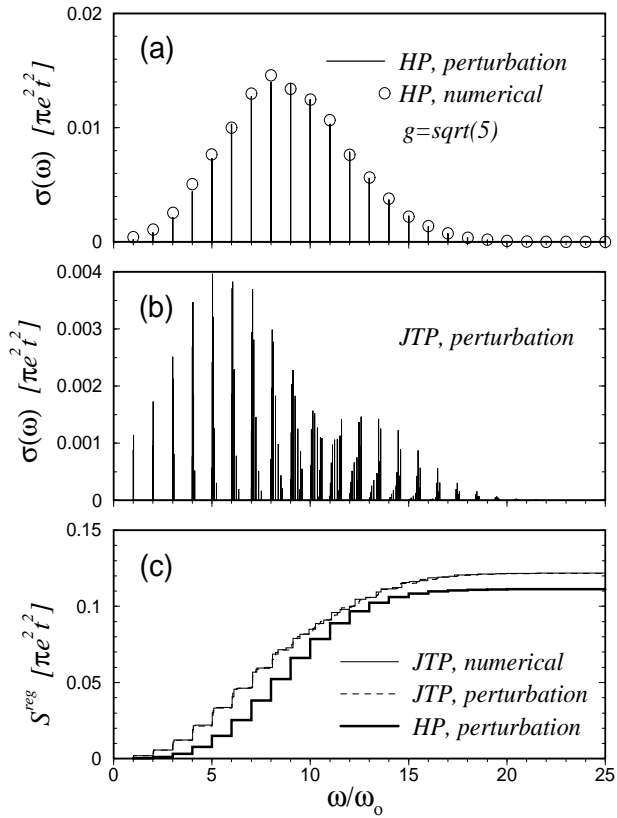,height=110mm, width=80mm,angle=0}
\end{center}
\caption{ Optical conductivity: second-order perturbation
theory is compared with the numerical result. The height of the peaks
represents the weight of the $\delta$-functions. The numerical calculations are
performed on the 2-site Holstein or Jahn-Teller models with $t=0.05$ and
$\omega_0=1$. The electron-phonon coupling is $g=\sqrt{5}$ for all panels.}
\label{fig:sigma_Li}
\end{figure}

For small hopping $t$, one can readily calculate optical conductivity
$\sigma^{reg}(\omega)$ using strong-coupling perturbation theory, i.e., the
Lang-Firsov transformation. For a two-site Holstein model \cite{N_site}, we
have
\begin{eqnarray}
\sigma^{reg}(\omega) & =& \sum_{n=1}^\infty 
\sigma_n \delta(\omega-n\omega_0) \nonumber \\
  & = & {\pi e^2 t^2} e^{-2g^2} \sum_{n=1}^\infty
\frac{(2g^2)^{n}}{n!\,\, n\ \omega_o} \delta (\omega  - n\omega _0)+ O(t^4).
\label{eq:HPsigma}
\end{eqnarray}
We note that $\sigma^{reg}(\omega)$ is not quite Poisson distributed.
It is composed of a series of Dirac $\delta$-
functions centered at the frequencies $\omega = n\omega_o + O(t)$. Their
weights, $\sigma_n$, are second-order in $t$. For $g \gg 1$, the largest
$\sigma_n $ in Eq.~\ref{eq:HPsigma} occurs when $n=2\omega_0(g^2-1) \sim 2E_p$,
which is consistent with previous numerical studies\cite{alex}.
Figure \ref{fig:sigma_Li}a shows good agreement between
perturbation theory and the numerical result.

In contrast, applying perturbation theory to the JTP 
is much more difficult because
the single-site problem does not have an exact solution. Reasonably accurate
trial wave functions for the low-lying levels have been
proposed\cite{Bersuker,Ku_JTP}.  However, obtaining the analytic wave 
functions for the
whole spectrum remains an arduous task. In the following, we evaluate the
second-order perturbation based on the numerically exact solutions of the
single-site problem.  It is known that the eigenfunctions $\phi_{n,j}$ of a JT
site are characterized by the radial quantum number $n$ and the angular quantum
number $j$. Each state is doubly degenerate and $n=0,~ j=\pm 1/2$ for ground
states. Thus, for a two-site JTP we chose the ground state to be
\begin{eqnarray}
|\psi_0\rangle = {1\over \sqrt{2} } \Big[ c_1^\dag |\phi_{n,1/2}\rangle |0,
0 \rangle + c_2^\dag |0, 0 \rangle  |\phi_{n,1/2}\rangle  \Big ]
\label{eq:JT_GS}
\end{eqnarray}
where $|0,0\rangle$ is the phonon state on the site without the electron.
We have
\begin{eqnarray}
\sigma^{reg} (\omega ) &=& \pi e^2 t^2 \sum_{n,m_1,m_2 =0}^\infty  \frac{{\left|
{\left\langle {\phi _n } \right.\left| {\left. {0,0} \right\rangle }
\right.\left\langle {m_1 ,m_2 \left| {\phi _0 } \right\rangle } \right.}
\right|^2 }}{{E_n  - E_0  + (m_1  + m_2 )\omega _0 }}\nonumber \\
 &\times&\delta [\omega  - E_n  +
E_0  - (m_1  + m_2 )\omega _0 ] + O(t^4), \label{eq:JTPT}
\end{eqnarray} where $E_n$ is the eigenvalue of $\phi_n$ and
$n + m_1 + m_2 > 0$. The quantum number
$j$ is left out because it is conserved by current
operator. All the nonzero matrix elements exist only between states
with $j=1/2$. In addition, the radial excitation of the JTP is anharmonic
due to its Mexican-hat potential surface. {\em Therefore, the
$\delta$-function peaks are not necessarily located near
$n\omega_0$}. We illustrate the distinct feature in
Fig.~\ref{fig:sigma_Li}b in which the anharmonicity manifests itself
clearly when the excited states come near the cone of the Mexican-hat
potential. (The height of the cone is $g^2\omega_0$ measured from the
bottom of the potential surface.) There are actually two peaks near
$\omega_0$ and three peaks near $2\omega_0$ because the excited phonon
quanta can either be harmonic or anharmonic. An excellent agreement
between perturbation theory Eq.~\ref{eq:JTPT} and Lanczos
diagonalization is found in Fig.~\ref{fig:sigma_Li}c. Although the
total weight of the JTP and HP are approximately equal, in the
low-frequency regime, we note that $S^{reg}_{JTP}(\omega) \gg
S^{reg}_{HP}(\omega)$ as a result of the fact that the JTP has a larger
quasiparticle weight than the HP\cite{takada,Ku_JTP}.

\begin{figure}[]
\begin{center}
\epsfig{file=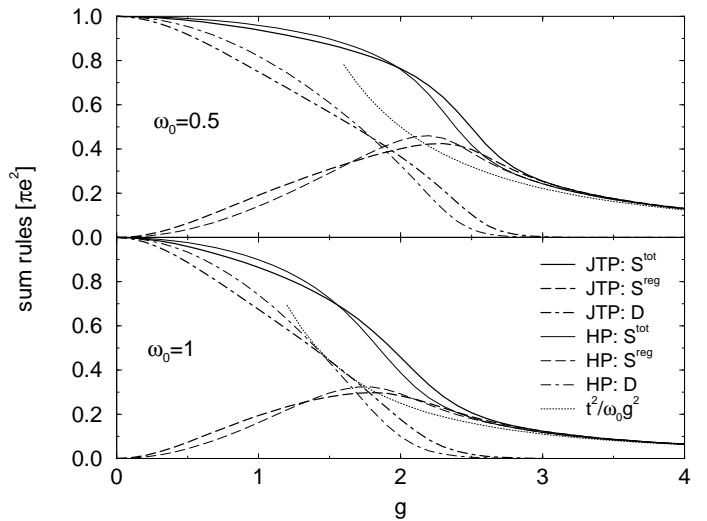,width=70mm,angle=-90}
\end{center}
\caption{Partial and total sum rules for $\omega_0=0.5$ (top) and
$\omega_0=1.0$ (bottom) as a function of coupling $g$, $t=1$. Thicker lines
represent JTP results, thinner lines HP.  The thin dotted line is a strong
coupling result.}
\label{fig:sumrule}
\end{figure}

Turning to the sum rules 
presented in
Fig.~(\ref{fig:sumrule}), we notice a monotonic decrease of the total
sum rule $S^{tot}$, which indicates a suppression of the electronic
kinetic energy in the strong coupling 
regime.  The drop in
$S^{tot}$ is accompanied by a decrease in the Drude weight $D$, which is
a measure of the coherent transport properties of a polaron. Although
the decrease is less steep in the JTP case, 
the two models show strong similarities.  The dependence of
the regular part $S^{reg}$ of the optical conductivity, which measures the
dissipative part of $\sigma(\omega)$, is not monotonic in the
electron phonon coupling $g$.
It first increases with increasing $g$, reaches the
peak near the crossover from large to small polaron regimes and then
decreases with further increase of $g$. Such an enhancement of
$S^{reg}$ in the intermediate coupling regime has already been
observed in the Holstein case \cite{fehske1}. In the strong coupling limit
$S^{tot}$ of {\bf both} models approach the asymptotic value
$S^{tot}\sim t^2/\omega_0g^2$ (see Fig.~(\ref{fig:sumrule})), which can
be obtained by integrating
Eq.~(\ref{eq:HPsigma}) and summing over all phonon excitations
in the limit $g \gg \omega_0$.
It is rather surprising that even though the strong coupling expansion is in
principle not valid for the JT case, $S^{tot}$ for the JT model seems
to approach the same asymptotic behavior. 
All quantities in Fig.~(\ref{fig:sumrule}) have
an error less than a linewidth.

\section{Bipolaron}

\begin{figure}[]
\begin{center}
\epsfig{file=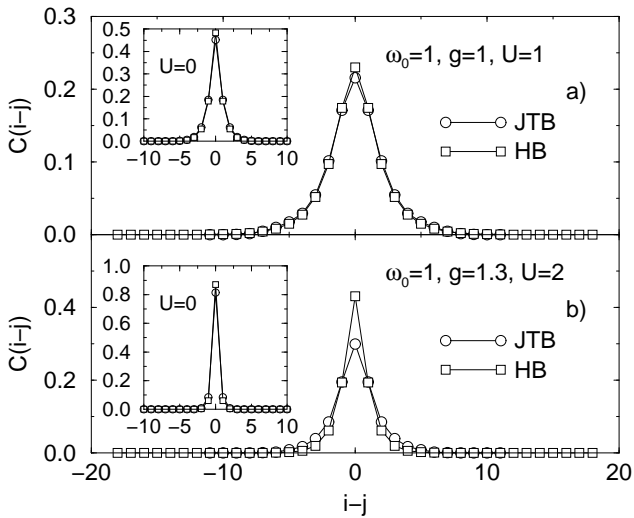,width=70mm,angle=-90}
\end{center}
\caption{$C(i-j)$ vs. $i-j$. We have used $N_h=18$ for HB and $N_h=11$
for JTB. }
\label{fig:elel}
\end{figure}

The following section does not apply to manganites, since
the strong Hund's rule coupling renders JT bipolarons unbound.
It does, however, apply to other transition metal oxides
with filled $t_{2g}$ shells, such as Ni or Co.
As we will see in this
section, similarities between the JT and Holstein models persist in the case of
two electrons. We start by comparing electron-electron correlation
function
\begin{equation} 
   C(i-j) = \innx{\psi_k}{n_i n_j}{\psi_k},
\label{eq:elel}
\end{equation}
where $ n_i = n_{i,1} + n_{i,2} $ and $\ket{\psi_k}$ is the bipolaron
wave function at momentum $k$. In Fig.~(\ref{fig:elel}a) and
Fig.~(\ref{fig:elel}b) we present $C(i-j)$ for 
$k=0$ and different values of the electron-phonon and Coulomb interactions.
For $U=0$ the JT bipolaron (JTB) and Holstein bipolaron (HB) show very
small differences. At smaller coupling $g=1$, Fig.~(\ref{fig:elel}a),
the effect of increased $U$ on the JTB and HB is similar. The main effect
of finite $U$ is to enlarge the size of the bipolaron, caused
by the electrons trying to avoid double occupancy. At slightly larger
coupling $g=1.3$, Fig.~(\ref{fig:elel}b), the effect of finite $U$ is
slightly more pronounced in the case of JTB than HB.  We can conclude
that the JTB in the intermediate to strong coupling regime has a slightly
larger radius at finite $U$ than the HB.

\begin{figure}[]
\begin{center}
\epsfig{file=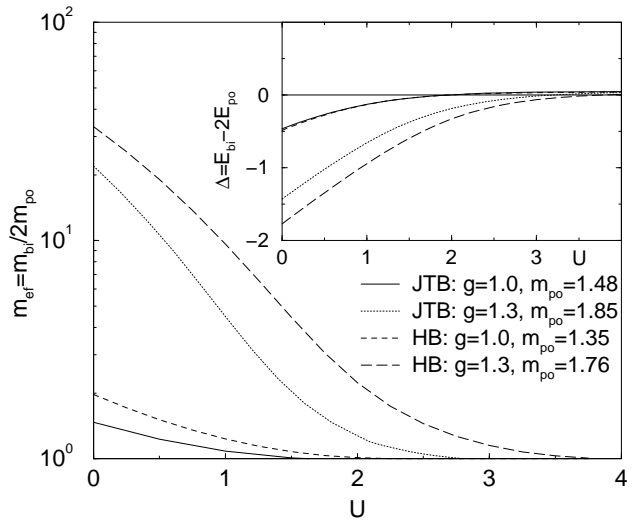,width=70mm,angle=-90}
\end{center}
\caption{Effective mass of JTB and HB vs. $U$ for
$\omega_0=1.0$. Inset: JTB and HB binding energies
vs. $U$. Finite-Hilbert space scaling was used for better accuracy. }
\label{fig:massbi}
\end{figure}

A larger radius of the bipolaron should imply
a larger mobility of the bipolaron. It is known that in the HB, with
increasing $U$ the on-site bipolaron transforms into an intra-site
bipolaron with a reduced effective mass
\cite{bonca1,magna,proville}. We may therefore expect the JTB to
have a smaller effective mass than the HB due to a larger radius
observed in Fig.~(\ref{fig:elel}b). This indeed turns out to be the
case as seen in Fig.~(\ref{fig:massbi}) where we show the bipolaron
effective mass vs. $U$ in units of the polaron mass for two different
coupling strengths. We see that JTB effective mass is smaller than HB
effective mass by up to a factor 2 for all $U$ and for both coupling
strengths. 
As shown in the inset of Fig.~(\ref{fig:massbi}), 
the JTB at $g=1.3$ has a smaller binding energy
than the HB, corresponding to its larger radius and smaller mass.
The binding energy is defined as 
$\Delta=E_{bi}-2E_{po}$, where $E_{bi}$ and $E_{po}$ are bipolaron and
polaron ground state energies respectively.
%

\begin{figure}[]
\begin{center}
\epsfig{file=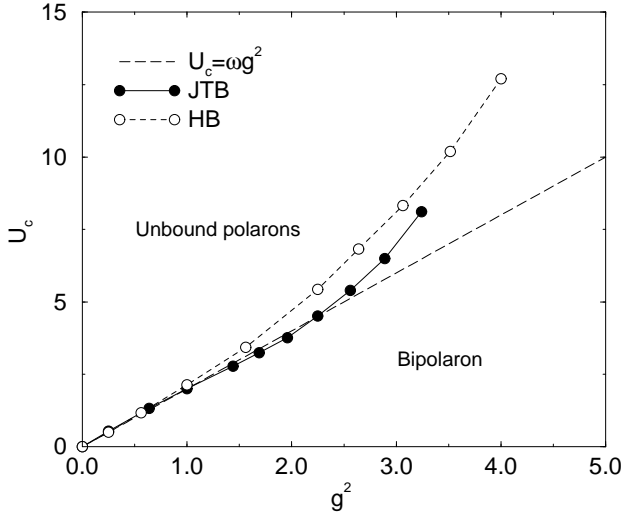,width=70mm,angle=-90}
\end{center}
\caption{The phase boundary between two separate polarons and
a bipolaron calculated for $\omega_0=1$. }
\label{fig:phd}
\end{figure}

We conclude our numerical investigation of the JTB with
the phase diagram presented in Fig.~(\ref{fig:phd}) for the transition
from unbound polarons to bound bipolarons. We have obtained the phase
diagram  using the condition
$\Delta(U,g)=0$.  It is important to stress that a bound bipolaron exists
in both models at $U=0$ and at any finite $g$. In the weak coupling
limit, {\it i.e. $g\ll 1$}, the phase boundary is given by
$U_c=2\omega g^2$, represented by the dashed line in
Fig.~(\ref{fig:phd}).  This result can be derived in the classical
limit and it is identical for both the JT and H  models. With increasing $g$, the
JTB phase boundary deviates slightly downwards from the given
analytical estimate, while for the HB it deviates upwards. Approaching
the strong coupling regime, $U_c(g)$ for the JTB seemingly approaches
phase boundary of the HB which in the strong coupling regime follows
$U_c=4\omega g^2$ \cite{bonca1}. Due to the large Hilbert space of the JTB
were unable to investigate the strong coupling regime of the JTB more
precisely in order to determine weather the two phase boundaries merge
with increasing $g$.

\section{Conclusions} 

  We have performed a detailed numerical analysis of the JTM in one
spatial dimension. The main conclusion is that the simple $E \otimes
e$ JTM is, despite extra orbital and phonon degrees of freedom,
very similar to the Holstein model (although the former has
a doubly degenerate ground state and the latter does not). 
This result agrees with recent
quantum Monte Carlo calculations \cite{korni} and analytical
comparison of both models \cite{takada}. Closer examination
reveals that the effective mass of the JTM is larger than HP effective
mass in the weak coupling regime and smaller by roughly a factor of
$g$ in the strong coupling regime, where the effective masses of JTP and HP
share the same exponent.

Turning to spectral properties, our numerical method gives extremely
reliable results for $\sigma^{reg}(\omega)$, which can be seen in weak
to intermediate coupling by observing the threshold frequency (the gap
in the spectra).  This agrees with the weak-coupling prediction, {\it
i.e.} $\omega_{tr}=\omega_0$. We also support this claim by providing
$\sigma^{reg}(\omega)$ for systems with different numbers of
variational states.  $\sigma^{reg}(\omega)$ at $g=1.0$ clearly shows
two broad peaks corresponding to one- and two- phonon emission.  These
peaks are much more pronounced in the JTM than in the HM. In the
strong-coupling regime, we find excellent agreement of
$\sigma^{reg}(\omega)$ for the Holstein model with a simple
Lang-Firsov strong coupling analytical expression. However, the
spectrum of the Jahn-Teller polaron is qualitatively different. Due to
the anharmonicity of its radial phonon excitation, the peaks of
$\sigma^{reg}(\omega)$ are not necessarily located near the multiples
of $\omega_0$.  For both models, the sum rule in this regime is given
by $S^{tot}\sim t^2/\omega_0 g^2$.


Strong similarities between HM and JTM persist also in the case of
two bound polarons. Similar to the one polaron case, the 
Jahn-Teller bipolaron has a smaller
effective mass than Hubbard bipolaron in intermediate coupling 
independent of
the strength of $U$. This is mostly a consequence of smaller binding
energy.

J.B.  gratefully  acknowledges the    support of Los   Alamos National
Laboratory   where part  of  this   work has  been   performed, the
financial support by the Slovene Ministry of Science, Education and Sport.
This work was supported in part by the US DOE and by LANL LDRD.


\begin{references}

\bibitem{millis} A.J. Millis, P.B. Littlewood, and Boris. I. Shraiman,
Phys. Rev. Lett. {\bf 74}, 5144 (1995).

\bibitem{zhao} G.-M. Zhao {\it et al}., Nature (London) {\bf 381}, 676
(1996).

\bibitem{quijada} M. Quijada, J. \v Cerne, J.R. Simpson, H.D. Drew,
K.H. Ahn, A.J. Millis,  S. Shreekala, R. Ramesh, M. Rajeswari, and
T. Venkatesan, Phys. Rev. B {\bf 58}, 16093 (1998). 

\bibitem{mareo} A. Moreo  {\it et al}., Science {\bf 283}, 2034
(1999).

\bibitem{brouet} V. Brouet, H. Alloul, Thien-Nga Le, S. Garaj, and
L. Forr\' o, Phys. Rev. Lett. {\bf 86}, 4680 (2001).

\bibitem{ole} O. Gunnarsson, Phys. Rev. B {\bf 51}, 3493 (1995).

\bibitem{auer} A. Auerbach, N. Manini, and E. Tosatti, Phys. Rev. B
{\bf 49}, 12998 (1994).

\bibitem{manini}  N. Manini, E. Tosatti, and A. Auerbach,
Phys. Rev. B {\bf 49}, 13008 (1994).

\bibitem{capone1} M. Capone, M. Fabrizio, P. Giannozzi, and
E. Tosatti, Phys. Rev. B {\bf 62}, 7619 (2000).



\bibitem{marsiglio} F. Marsiglio, Physica C {\bf 244}, 21 (1995).

\bibitem{wellein} G. Wellein, H. R\" oder, and H. Fehske, Phys. Rev. B
{\bf 53}, 9666 (1996); G. Wellein and H. Fehske, Phys. Rev. B {\bf
56}, 4513 (1997); Wellein and H. Fehske, Phys. Rev. B {\bf 58}, 6208
(1998); A. Weise, H. Fehske, G. Wellein, and A.R. Bishop, Phys. Rev. B
{\bf 62}, R747 (2000). 

\bibitem{mello} E. V. L. de Mello and J. Ranninger, Phys. Rev. B {\bf
55}, 14872 (1997).

\bibitem{bonca}  J. Bon\v ca, S. A. Trugman and I. Batisti\v c,
Phys. Rev. B {\bf 60}, 1633 (1999).

\bibitem{lichung} Li-Chung Ku, S.A. Trugman, and J. Bon\v ca,
Phys. Rev. B {\bf 65}, 174306 (2002).

\bibitem{raedt} H. De Raedt and A. Langedijk, Phys. Rev. Lett {\bf
49}, 1522 (1982); H. De Raedt and A. Langedijk, Phys. Rev. B {\bf 27},
6097 (1983); {\bf 30}, 1671 (1984).

\bibitem{kornilovitch} P. E. Kornilovitch and E. R. Pike, Phys. Rev. B
{\bf 55}, R8634 (1997).

\bibitem{romero} A. W. Romero, D. W. Brown and K. Lindenberg,
J. Chem. Phys. {\bf 109}, 6540 (1998).

\bibitem{white} E. Jeckelmann and S. R. White, Phys. Rev. B 
{\bf 57}, 6375 (1998).



\bibitem{fehske1} H.\ Fehske, J.\ Loos, and G.\ Wellein, Z. Phys. B
{\bf 104} , 619 (1997).

\bibitem{capone} M. Capone, W. Stephan, and M. Grilli, Phys. Rev. B
{\bf 56}, 4484 (1997).

\bibitem{fehske} H. Fehske, J. Loos, and G. Wellein, Phys. Rev. B {\bf
61}, 8016 (2000).

\bibitem{alex} A.S. Alexandrov, V.V. Kabanov, and D.K. Ray, Physica C
{\bf 224}, 247 (1994).

\bibitem{zhang} Chunli Zhang, E. Jeckelmann, and S.R. White,
Phys. Rev. B {\bf 60}, 14092 (1999). 


\bibitem{emin} D. Emin, Phys. Rev. B {\bf 48}, 13691 (1993).

\bibitem{lowfreq}  At strong coupling, the Jahn-Teller polaron
has phonon excitations at much lower frequency than $\omega _ 0$
(angular excitations of the Mexican hat),
but these cannot be accessed optically for the type of hopping
considered here.



\bibitem{korni} P.E. Kornilovitch, Phys. Rev. Lett. 84, 1551 (2000).



\bibitem{bonca1}  J. Bon\v ca, T. Katra\v snik, and S. A. Trugman,
Phys. Rev. Lett. {\bf 84}, 3153 (2000).

\bibitem{magna} A. La Magna and R. Pucci, Phys. Rev. B {\bf 55}, 
14886 (1997).

\bibitem{proville} L. Proville and S. Aubry, Physica D {\bf 133}, 307
(1998); L. Proville and S. Aubry, preprint.

\bibitem{takada} Y. Takada,  Phys. Rev. B 61, 8631 (2000).


\bibitem{N_site} It can be shown that for N sites ($N > 2$) with periodic 
boundary conditions, the result of the perturbation is approximately 
a factor of two larger than that of the two-site model.

\bibitem{Bersuker} I.~B.~Bersuker and V.~Z.~Polinger, {\it Vibronic
Interactions in Molecules and Crystals} (Springer, Berlin, 1989).
\bibitem{Ku_JTP} L.-C.~Ku, S.~A.~Trugman, to be published.










\end{references}
\end{document}